\documentclass[12pt]{article}

\pdfoutput=1

\usepackage{amsmath,amssymb,amsfonts,amscd,mathrsfs,bm,mathtools}
\usepackage[dvipsnames]{xcolor}
\definecolor{darkblue}{rgb}{0.1,0.1,.7}
\usepackage[colorlinks, linkcolor=darkblue, citecolor=darkblue, urlcolor=darkblue, linktocpage]{hyperref} 
\usepackage[square, comma, compress,numbers]{natbib}
\usepackage[]{graphicx}
\usepackage{geometry}
\usepackage[margin=10pt,font=small,labelfont=bf]{caption}
\usepackage{ifthen}
\usepackage{slashed}

\usepackage{cancel}

\usepackage{subcaption}
\usepackage{booktabs,multirow}
\usepackage{hhline}
\usepackage{tablefootnote}

\usepackage{tablefootnote,colortbl}
\definecolor{Gray1}{gray}{0.97}
\definecolor{Gray2}{gray}{0.9}
\definecolor{LightCyan}{rgb}{0.88,1,1}
\definecolor{blu}{rgb}{0,0,1}

\usepackage{ragged2e}
\usepackage{array}
\newcolumntype{L}[1]{>{\raggedright\let\newline\\\arraybackslash\hspace{0pt}}m{#1}}
\newcolumntype{C}[1]{>{\centering\let\newline\\\arraybackslash\hspace{0pt}}m{#1}}
\newcolumntype{R}[1]{>{\raggedleft\let\newline\\\arraybackslash\hspace{0pt}}m{#1}}

\usepackage{booktabs}

\usepackage{titlesec}
\titleformat*{\section}{\large\bfseries}
\titleformat*{\subsection}{\normalsize\bfseries}
\titleformat*{\subsubsection}{\normalsize\it}
\titleformat*{\paragraph}{\normalsize\bfseries}
\titleformat*{\subparagraph}{\normalsize\bfseries}

\newcommand{\beq}{\begin{equation}} 
\newcommand{\eeq}{\end{equation}}

\newcommand{\diffop}[2]{\ifthenelse{\equal{#2}{1}}{\frac{\mrm{d}}{\mrm{d} #1}}{\frac{\mrm{d}^#2}{\mrm{d} #1^#2}}}

\newcommand{\mrm}[1]{{\mathrm #1}}

\usepackage[normalem]{ulem}

\catcode`,\active

\catcode`\,12

\newcommand{\be}{\begin{equation}}
\newcommand{\ee}{\end{equation}}
\newcommand{\bea}{\begin{eqnarray}}
\newcommand{\eea}{\end{eqnarray}}
\newcommand{\eq}[1]{Eq.~(\ref{#1})}

\usepackage{accents}
\newlength{\dhatheight}

\newcommand{\br}[1]{\text{Br}}

\numberwithin{equation}{section}
\usepackage{setspace}
\newcommand{\la}{\langle}
\newcommand{\ra}{\rangle}

\begin{document}

\begin{titlepage}
\begin{center}
\hfill

\vspace{2.0cm}
{\Large\bf  
Wilson Coefficients and Natural Zeros \\
\vspace{0.5cm}
from the  On-Shell Viewpoint}
\vspace{2.0cm}
\\
{\bf 
Luigi Delle Rose$^{a}$,
Benedict von Harling$^{a}$
 and
Alex Pomarol$^{a,b,c}$
}
\\
\vspace{0.7cm}
{\it\footnotesize
${}^a$IFAE and BIST, Universitat Aut\`onoma de Barcelona, 08193 Bellaterra, Barcelona\\
${}^b$Departament de F\'isica, Universitat Aut\`onoma de Barcelona, 08193 Bellaterra, Barcelona\\
${}^c$CERN, Theory Division, Geneva, Switzerland
}

\vspace{0.9cm}
\abstract
We show how to simplify the calculation of  the finite  contributions  
 from heavy particles to EFT Wilson coefficients   by  using on-shell methods. 
We apply the technique to  the one-loop calculation  of  $g-2$ and $H\gamma\gamma$,
 showing how finite contributions can be obtained from the product of tree-level amplitudes.
In certain cases, due to a parity  symmetry of these amplitudes,  the total contribution adds up to
zero, as previously found in the literature. Our method allows to search for new natural zeros,
as well as  to obtain non-zero contributions in a straightforward way.
\end{center}
\end{titlepage}
\setcounter{footnote}{0}


\section{Introduction}

The Effective Field Theory (EFT)  approach to describe low-energy experiments is based on the idea of 
integrating out heavy particles from the theory,  keeping only the relevant light degrees of freedom.
The effects of  these heavy particles are  then captured in the (Wilson) coefficients of the EFT
which become the  relevant parameters to be measured in low-energy experiments, as for example the
magnetic dipole moment of the SM fermions, $g-2$.

When performing these calculations  using Feynman rules,
it has been found that  in certain models 
the different contributions add up to zero with no apparent explanation, jeopardizing the idea of naturalness
which states  that {\it contributions not forbidden by symmetries are  compulsory}.
 An example for the $g-2$ case has been  extensively studied in \cite{Arkani-Hamed:2021xlp},
and other examples can be found e.g.~in  \cite{Panico:2018hal}
for the Higgs coupling to photons, $H\gamma\gamma$.

Here we will analyze  these cancellations  by calculating the finite contributions  to  the Wilson coefficients
 using on-shell methods. 
 These methods have already been useful to understand certain cancellations  in the anomalous dimensions
which  looked mysterious from the  Feynman approach \cite{Cheung:2015aba}.
We will show that the finite contributions 
to $g-2$ arising from integrating out heavy fermions
can also be obtained by a product of tree-level amplitudes. 
In certain models, these amplitudes are odd under the interchange of the heavy fermions,
while the total contribution must be even under this interchange. 
This will explain the vanishing of the total contribution in the models of \cite{Arkani-Hamed:2021xlp}, and 
will allow to find new cases where the contributions also add up to zero.
For models without this parity, the contributions will not cancel, and our method will explicitly provide 
the  Wilson coefficients as  a simple product of tree-level amplitudes.

We will extend the analysis also to the Wilson coefficient of the $H\gamma\gamma$ coupling 
(for tree-level calculations using on-shell methods, see  \cite{Shadmi:2018xan}).
We will see that the same argument as for the $g-2$ case can lead to an explanation for the 
absence of the total contribution to this Wilson coefficient
in certain models of heavy fermions.

The calculation of  Wilson coefficients  using on-shell amplitudes
has been previously studied in the literature
 - see for example \cite{Bern:2021ppb} and references therein.
Nevertheless, in these cases first the full amplitude is calculated,  and later
the heavy mass limit is taken to match with the EFT.
This makes the method too long and probably not so competitive with the Feynman approach.
The main purpose here will be to understand what cuts in the one-loop amplitudes are needed   in order 
to simply  extract the finite contributions to the Wilson coefficients,  specially for  cases of phenomenological interest. 
This will also allow us to better understand the origin of the rational number appearing in the  Wilson coefficients.

While this work was being written, the article \cite{Craig:2021ksw}
 appeared in the archives where 
also a  symmetry argument was presented as an explanation of the zeros found in \cite{Arkani-Hamed:2021xlp}.
Although the symmetry is also an interchange parity, the approach in  \cite{Craig:2021ksw} is different from the one followed here~\cite{conf}.

\section{Finite contributions to Wilson coefficients via on-shell methods}

Amplitudes at the one-loop level  can have a Passarino-Veltman decomposition given by
\be
{\cal A}_{\rm loop}=\sum_a C_1^{(a)} I_1^{(a)}+\sum_b C_2^{(b)} I_2^{(b)}+\sum_c C_3^{(c)} I_3^{(c)}+\sum_d C_4^{(d)} I_4^{(d)}+R\,,
\label{general}
\ee
where $I_n$  are master scalar integrals with $n$ propagators ($n=1,2,3,4$) and $C_n$ are mass- and kinematic-dependent coefficients. The master integrals are given by
\be
I_n=(-1)^n\mu^{4-D} \int\frac{d^D\ell}{i(2\pi)^D}\frac{1}{(\ell^2 - M^2_0) \, ((\ell-P_1)^2 - M^2_1) \, ((\ell-P_1-P_2)^2 - M^2_2)\cdots}\,,
\ee
where  $P_1,P_2,...,P_{n-1}$ are sums of external particle momenta $p_i$.
The first four contributions to \eq{general} are called respectively tadpoles, bubbles, triangles and 
boxes, according to the topology of the scalar integral.  
Terms collected under $R$ are rational functions of  kinematic invariants. 
We will be using dimensional regularization, $D=4-2\epsilon$.

In general ${\cal A}_{\rm loop}$ can be divergent. Nevertheless, here we are only interested in
one-loop effects from renormalizable theories contributing to processes, such as the magnetic dipole moment, which must go to zero as the heavy masses go to infinity, and are therefore UV convergent.  
Even in these cases, it is still possible to have IR divergencies ($\propto 1/M_i^2 \ln M_i/\mu$)  which would signal 
the presence of nonzero anomalous dimensions.  The processes that we will consider here will however also be IR convergent.\footnote{For obtaining anomalous dimensions via on-shell methods, see for example 
\cite{Caron-Huot:2016cwu,2,Baratella:2020lzz,4,5,6,7,8}.}

To match with the EFT, we must take the 
limit in which the masses of the  heavy particles in ${\cal A}_{\rm loop}$  are larger than all the external momenta $p_i$.
For simplicity in this section we take these masses to be equal to $M$. Note that $I_n$ can involve both massless,
$M_i=0$, as well as massive states, $M_j=M$. The external states will be assumed to be massless, $p^2_i=0$.
By performing an expansion for $P_{i}\ll M$ in \eq{general},
the amplitude ${\cal A}_{\rm loop}$ should match with the amplitude associated with a higher-dimensional  operator 
${\cal A}_{{\cal O}_i}=\langle 12...|{\cal O}_i|0\rangle$.
Assuming that the leading operator in this expansion has 
dimension six, we have 
\be
{\cal A}_{\rm loop}\to \frac{C_i}{M^2} {\cal A}_{{\cal O}_i}+\cdots\,,
\ee
where $C_i$ is a rational number times some couplings divided by $16\pi^2$, and is often referred to as the Wilson coefficient of the corresponding operator.
In Appendix~\ref{appendixa} we will prove 
that no contribution at order $1/M^2$ can arise from $R$ in \eq{general}.
Therefore we have that the Wilson coefficients are given by
\be
C_i= \frac{1}{\cal A}_{{\cal O}_i}\, \lim_{P_i/M\to 0} M^2\left(\sum_a C_2^{(a)} I_2^{(a)}+\sum_b C_3^{(b)} I_3^{(b)}+\sum_c C_4^{(c)} I_4^{(c)}\right)\,.
\label{master}
\ee 
From \eq{master} we  see that in order to determine the Wilson coefficients
 we need to know the coefficients $C^{(a)}_{n}$.
These coefficients can however be easily obtained using on-shell methods.
In particular
generalized unitarity methods, extensively developed in the literature in recent years \cite{Dixon:2013uaa}, allow one to calculate $C^{(a)}_{n}$  without the need to perform loop calculations. Instead one uses products of tree-level amplitudes (integrated over some phase space), making the determination of the Wilson coefficients $C_i$ clearer.
The idea is to obtain  the coefficients $C^{(a)}_n$ from performing  $n$-cuts in the loop. Although this can look like  a lengthy procedure, we will see that in many  cases, and specially those we are interested in, the situation is quite simple and only 
one or two  2-cuts are needed.

\section{Dipole moment Wilson coefficient}

We start by   calculating the  Wilson coefficient of the magnetic dipole moment   induced by different models with heavy fermions. 
For the SM leptons this operator is defined as 
\be
\frac{C_\gamma}{2M^2}\frac{q_e}{\sqrt{2}}\,  \bar \ell_L\sigma_{\mu\nu}e_R H  F^{\mu\nu}=
\frac{C_\gamma}{M^2}\, q_e\, \ell_\alpha e_\beta H F^{\alpha\beta}
\,,
\label{dipole}
\ee
where we have introduced the 2-component Weyl spinors $\ell_\alpha$ and $e_\alpha$ of helicity $h=-1/2$,
and $F$ refers to  the field strength of the photon.  We follow the usual definition of the gauge coupling 
used when calculating amplitudes with spinors ($\Delta {\cal L}=q_f A_\mu\bar f\gamma^\mu f/\sqrt{2}$)
which avoids the proliferation of factors of $\sqrt{2}$ in the calculations.
In particular, \eq{dipole}  leads  to the amplitude  
\be
\frac{C_\gamma }{M^2} {\cal A}_{D}(1_{\ell},2_{e}, 3_{\gamma_-},4_{H^0})
=\frac{C_\gamma}{M^2}\, q_e\,\la 13\ra \la 23\ra\,,
\ee
where  $H^0$ is the neutral Higgs component, and with an abuse of notation we have denoted by $\ell$ also 
the  charged-lepton component of the SU(2)$_L$ doublet. Subindices $\pm$ denote helicities $h=\pm1$ and 
all particles are taken to be  incoming.
We use spinor-helicity notation~\cite{Dixon:2013uaa} using  properties and conventions which are summarized in Appendix~\ref{appendixb}.
A key point for our calculation is to  set the Higgs momentum to zero, $p_{H^0}=0$,
as this   enormously simplifies the  loop amplitude.

\subsection{Massive vector-like singlet $S$ and doublet $L$}
\label{SLmodel}

The first model we consider is the one studied in Ref.~\cite{Arkani-Hamed:2021xlp} which consists of two massive vector-like fermions, a singlet ($S$) and an SU(2)$_L$ doublet ($L$). The Lagrangian in Weyl notation  is given by (omitting Lorentz indices)
\be
{\cal L}=-Y_L  \ell S  \tilde H-Y_R  L e  H-
Y_{V}  \tilde  H^\dagger L^c S^c - Y'_{V}    LS \tilde H -M_S\, S S^c- M_L\,   L L^c+h.c.\,,
\label{modelSL}
\ee
where $\tilde H=i\sigma_2H^*$. For simplicity we choose the couplings to be real.
As we will see, it will be useful for our  calculation to take  $M_S\not=M_L$ with  both being larger than the $P_i$.
As in Ref.~\cite{Arkani-Hamed:2021xlp},  we set the SM Yukawa coupling for the muon to zero, $Y_\ell=0$ (at tree-level).

\begin{figure}[t]
\centering
\includegraphics[width=0.35\textwidth]{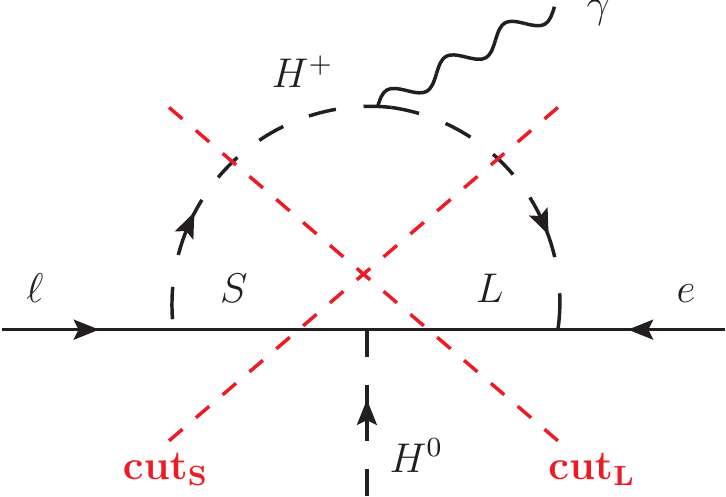}
\caption{\it One-loop   contribution to the $g-2$ of the SM leptons
from the model \eq{modelSL} with $Y_V=0$ and $Y'_V\neq 0$, with the  relevant 2-cuts.}
\label{dipoleSL}
\end{figure}

\noindent{\bf $\bullet \ Y'_V\not=0$ case:}
Let us start by considering the case  with $Y_V=0,\,  Y'_V \neq0$.
 The Feynman diagram which contributes to the SM lepton dipole moment is given in Fig.~\ref{dipoleSL}.
We will follow \eq{master} for the calculation where the coefficients $C^{(a)}_n$ will be obtained via on-shell amplitudes
from  $n$-cuts.

The first thing to realize in this  example, and the others that will follow, 
is  that all the coefficients $C_4$ and $C_3$ vanish, as all possible 4-cuts and  3-cuts of  Fig.~\ref{dipoleSL} give zero.
The reason is the following.
Since we are taking $p_{H^0}=0$,  we have $p_S=p_L$ and then the condition to have $S$ and $L$ simultaneously on-shell cannot be fulfilled as both have different masses. 
This implies that the 4-cut is zero (no boxes) and the only potential nonzero 3-cut  
must arise from  cutting  two massless states and one massive state.
This corresponding triangle is  however  also zero.
Indeed, one can follow the 
arguments of  Ref.~\cite{Baratella:2020lzz} to prove that  in the absence of IR divergencies (as it is our case),
IR-divergent triangles cannot be  present when there are no boxes.
We are then left  only with  bubbles. 

We  can obtain the bubble coefficients $C_2$ from 2-cuts. 
Before performing calculations it is important to remark that 
since $p_{H^0}=0$, we have $p_1+p_2+p_3=0$.
Therefore we can work  either in the limit in which $s_{13}=(p_1+p_3)^2$ is small but nonzero, but then we will have to take also
$p^2_2=s_{13}\not=0$ (i.e.~the fermion $e$  slightly off-shell), or
alternatively, we can take the limit $s_{23}=(p_2+p_3)^2\to 0$ and then 
$p^2_1=s_{23}\not=0$ (i.e.~the fermion $\ell$  slightly off-shell).
Let us choose the first option and consider the 2-cuts  where $S$ becomes on-shell.
There are in principle two possible 2-cuts of this type. However,  the one leaving $\ell$ alone as an external leg is proportional to 
$I_2(p^2_1=0,M^2_S,0)$ and cannot  give any contribution of $O(s_{13}/M^2)$.\footnote{
 All these types of bubbles, added to the   $s_{13}$-independent terms of the bubble \eq{BubbleFunctionExpansion},
   must sum to zero since the one-loop amplitude cannot have divergent terms.}
The only relevant 2-cut is then the one depicted by ${\bf cut_{\bf S}}$
in Fig.~\ref{dipoleSL}.
We  have
 \be
C^{(13)}_2=
\int d{\rm LIPS}\,  (-1)^F {\cal A}(1_\ell,3_{\gamma_-},1'_{S},3'_{H^+})\times {\cal A}(3'_{\bar H^+},1'_{\bar S},2_e,4_{H^0})\,,
\label{C2}
\ee
where the  integral is over the   Lorentz-Invariant Phase Space (LIPS) associated with  the momenta
of the  two cut states, $p_{1'}$ and  $p_{3'}$, normalized as $\int d{\rm LIPS}=1$.
With a bar over a state we denote that the signs of the  momentum, helicity and all other quantum numbers of the state
have been reversed, and $F$ is the number of internal fermions ($F=1$ in this case) \cite{Baratella:2020lzz}.

The tree-level amplitudes in \eq{C2}  can be easily calculated from the model \eq{modelSL}.
We use the spinor-helicity formalism for massive particles from Ref.~\cite{Arkani-Hamed:2017jhn}, using  properties and conventions which are summarized in Appendix~\ref{appendixb}. This gives (recall that $p_{H^0}=0$)
\be
{\cal A}(1_\ell,3_{\gamma_-},1'_{S},3'_{H^+})=q_e Y_L M_S\,\frac{[3'{\bf1'}]}{[3'3][13]}\,, \ \ \ \ \ \ \ 
{\cal A}(3'_{\bar H^+},1'_{\bar S},2_e,4_{H^0})=Y_RY'_{V}\frac{[-{\bf1'}|p_{1'}|2 \ra}{M^2_S-M_L^2}\,.
\label{amplitudesModelSL}
\ee
Writing the SU(2) little-group indices of the bold spinor-helicity variables explicitly, the integrand in \eq{C2} is then given by
\bea
&& {\cal A}(1_\ell,3_{\gamma_-},1'^I_{S},3'_{H^+}) \, \epsilon_{IJ} \, 
{\cal A}(3'_{\bar H^+},1'^J_{\bar S},2_e,4_{H^0}) \, \, = - q_eY_LY_RY'_V
\frac{M_S^2}{M^2_S-M_L^2}\frac{[3'|(p_3+p_1)|2\ra}{[3'3][13]}\nonumber\\
&&= - q_eY_LY_RY'_V\frac{M_S^2}{M^2_S-M_L^2}\left(\frac{\la 32 \ra}{[13]}+
\frac{[3'1]\la12\ra}{[3'3][13]}\right),
\label{AA}
\eea
where 
we have used that $|1']^I[-1'|_I= M_S$ (see Appendix~\ref{appendixb}).
Only the first term can give a contribution to the dipole since in the second term the  two external spinors are contracted among themselves, $\la 12\ra$.\footnote{The vanishing of the second term can also be explicitly seen by  integrating over 
the phase space as  in \eq{C2}  which can be easily done by  relating the  internal spinor $| 3' ]$ with the external
ones $| 1]$ and $| 3]$.  This relation is given by
$| 3' ]= \sqrt{1 - M_S^2/s_{12}} \left(  c_{\theta/2}  | 1 ] + s_{\theta/2} e^{-i \phi}  | 3 ] \right)$
which fulfills the kinematic constraint $p_1+p_3=p_{3'}+p_{1'}$ with
$p_1^2=p_3^2=p_{3'}^2=0$ and $p_{1'}^2=M_S^2$. The integral measure is
$\int {\rm dLIPS}= \int^{2\pi}_0 d \phi \int_0^\pi d \theta\, s_\theta/(4\pi)$.
}
The first term of \eq{AA} does not depend on the internal spinors so the dLIPS integration in \eq{C2} is trivial 
leading  to 
\be
C^{(13)}_2
=Y_LY_RY'_V \frac{M_S^2}{M^2_S-M_L^2}\frac{1}{s_{13}} {\cal A}_{D}(1_{l},2_{e}, 3_{\gamma_-},4_{H^0})\,.
\ee
Plugging this into  \eq{master},
and expanding the bubble integral to $O(s_{13}/M^2_S)$,
\be
\label{BubbleFunctionExpansion}
I^{(13)}_2(s_{13},M_S^2,0)\, \simeq \, \frac{1}{16 \pi^2} \left( \frac{1}{\epsilon} + \ln \frac{\mu^2}{M^2_S} + 1 +\frac{s_{13}}{2M^2_S}+\cdots \right)\,,
\ee
we obtain  a finite  term\footnote{Again, we are neglecting the divergent and the constant term in 
\eq{BubbleFunctionExpansion} which must cancel other divergent and $1/s_{13}$ contributions 
arising from  bubbles proportional to $I_2(0,M_S^2,0)$ which, as explained, we also neglect.}
corresponding to the contribution from this 2-cut to the Wilson coefficient:
\be
\frac{\Delta C_{\gamma}}{M^2}
=\frac{Y_LY_RY'_V}{32\pi^2}\frac{1}{M_S^2-M^2_L}\,.
\label{WilsoncutS}
\ee

There are also 2-cuts where  $L$ instead of $S$ is put on-shell,
in particular ${\bf cut_{\bf L}}$ of Fig.~\ref{dipoleSL}.
It is clear however that this 2-cut is identical to ${\bf cut_{\bf S}}$  by the exchange\footnote{If we keep $s_{13}\not=0$
the contribution from this 2-cut will not contain terms of $O(s_{ij}/M^2_L)$ and can be neglected.
However,  there is then another 2-cut which isolates the fermion $e$ with $p^2_2\not=0$ and which does not vanish. 
Since the  contribution from cutting the $L$ state
should not depend on our choice of whether we take $s_{12}$ or $s_{13}$ nonzero,
we can just choose $s_{12}\not=0$ for this calculation in which case ${\bf cut_{\bf L}}$ gives  the only relevant contribution.}
\be
S\leftrightarrow L\,, \ \ \ \ \
\ell \leftrightarrow e\,.
\label{sym}
\ee
Therefore the contribution must be the same as \eq{WilsoncutS} with the replacement $M_S\leftrightarrow M_L$.
Since \eq{C2contS} is odd under this transformation,  the total contribution to $C_\gamma$  adds up to zero.

It is easy to  understand this cancellation  without the need  to go through all the details of the calculation.
The first important   thing to know is how $M_L$ enters into ${\bf cut_{\bf S}}$, 
since the  dependence on  $M_S$ can then be  fixed by dimensional analysis.
Now, by inspection of  the second amplitude of \eq{amplitudesModelSL} we see that, due to the $L$ propagator, 
$M_L$  can only appear as $\Delta C_\gamma\propto 1/(M_S^2-M_L^2)$, being then odd under \eq{sym}.
Since the total contribution from   ${\bf cut_{\bf S}}$ and ${\bf cut_{\bf L}}$
must be symmetric under \eq{sym}, this must be zero.

\begin{figure}[t]
\centering
\includegraphics[width=0.8\textwidth]{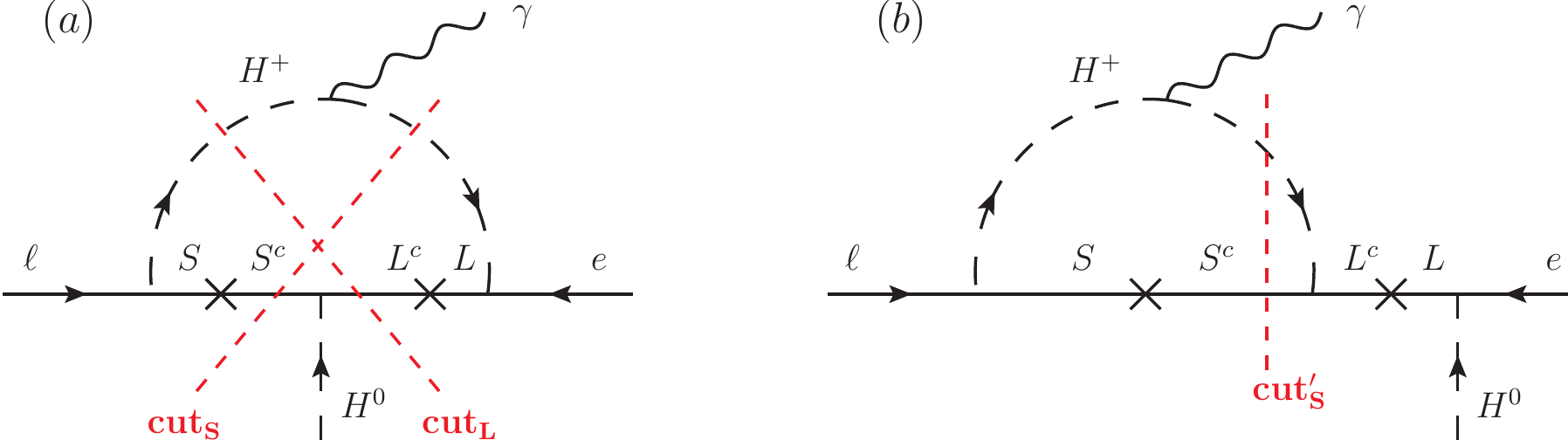}
\caption{\it 
One-loop   contributions to the $g-2$ of the SM leptons
from the model \eq{modelSL} with $Y'_V=0$ and $Y_V\neq 0$, with the relevant 2-cuts.}
\label{dipoleSLb}
\end{figure}

\noindent{\bf  $\bullet \ Y_V\not=0$ case:}
Let us next move to  the case $Y'_V=0,\,  Y_V \neq0$. 
We now have two Feynman diagrams, see Fig.~\ref{dipoleSLb}. 
Let us first consider diagram (a).
The contribution from  ${\bf cut_{\bf S}}$  is again given by \eq{C2}
  with the  only difference with respect to the previous case that now
\be
{\cal A}(3'_{\bar H^+},1'_{\bar S},2_e,4_{H^0})=Y_RY_{V}{\la-{\bf1'} 2 \ra}\frac{M_L}{M^2_S-M_L^2}\,.
\label{HHSe}
\ee
Plugging \eq{HHSe}  into \eq{C2} it is easy to see that we get the same as in \eq{WilsoncutS}
 with the only differences being that, due to the $L$ mass insertion in \eq{HHSe}, we have 
an extra factor of $M_L/M_S$ and that it depends on $Y_V$ instead of $Y'_V$:
\be
\frac{\Delta C_{\gamma}}{M^2}
=\frac{Y_LY_RY_V}{{32}\pi^2}\frac{M_L/M_S}{M_S^2-M^2_L}\,.
\label{WilsoncutS2}
\ee
This contribution is not odd under the symmetry  \eq{sym}. Therefore, when adding ${\bf cut_{\bf L}}$ of Fig.~\ref{dipoleSLb}, obtained by performing $M_S\leftrightarrow M_L$ in \eq{WilsoncutS2}, we get a nonzero result:
\be
\frac{\Delta C_{\gamma}}{M^2}
=\frac{Y_LY_RY_V}{{32}\pi^2}\frac{(M_L/M_S-M_S/M_L)}{M_S^2-M^2_L}= -\frac{Y_LY_RY_V}{{32}\pi^2}\frac{1}{M_SM_L}\,.
\label{WilsoncutS3}
\ee
There is, however, another contribution coming from the 2-cut  of diagram (b) in Fig.~\ref{dipoleSLb}.
This contribution can be considered as arising from an extra term to  the amplitude \eq{HHSe} given by
\be
 {\cal A}(3'_{\bar H^+},1'_{\bar S},2_e,4_{H^0}) \, = \, -Y_R Y_{V}{\la-{\bf1'} 2 \ra}\frac{M_L}{p_2^2-M_L^2} \, \simeq \, 
 Y_R Y_{V}{\la-{\bf1'} 2 \ra}\frac{1}{M_L}\,,
\label{HHSeextra}
\ee
which exactly cancels the leading term of $O(1/M_L)$ of this amplitude in the limit  $M_L\gg M_S$. 
Notice that this leading term was crucial in obtaining the nonzero result at $O(1/M_L M_S)$ in \eq{WilsoncutS3}. We therefore have that by adding the contribution from \eq{HHSeextra} in \eq{C2} we again find a vanishing Wilson coefficient.

The origin of this cancellation can again be  understood from symmetries.
For $M_L\gg M_S$, the leading term 
of the total amplitude  ${\cal A}(3'_{\bar H^+},1'_{\bar S},2_e,4_{H^0})$
is captured by the dimension-5 operator
$\tilde H^\dagger H S^ce/M_L$. However,  this operator  is zero since $\tilde H^\dagger H=\epsilon_{ab}H^aH^b=0$ 
($a,b$ being SU(2)$_L$ indices).
Were  this property  absent, as we will find in the next model,
the Wilson coefficient would have been generated.

We conclude then that the {\it a priori} non-trivial result that the contributions to the 
Wilson coefficient of the dipole moment add up to zero in this model boils down, by inspection with on-shell methods, to
a clash of $even\times odd$ under a given parity.

\begin{figure}[t]
\centering
\includegraphics[width=0.3\textwidth]{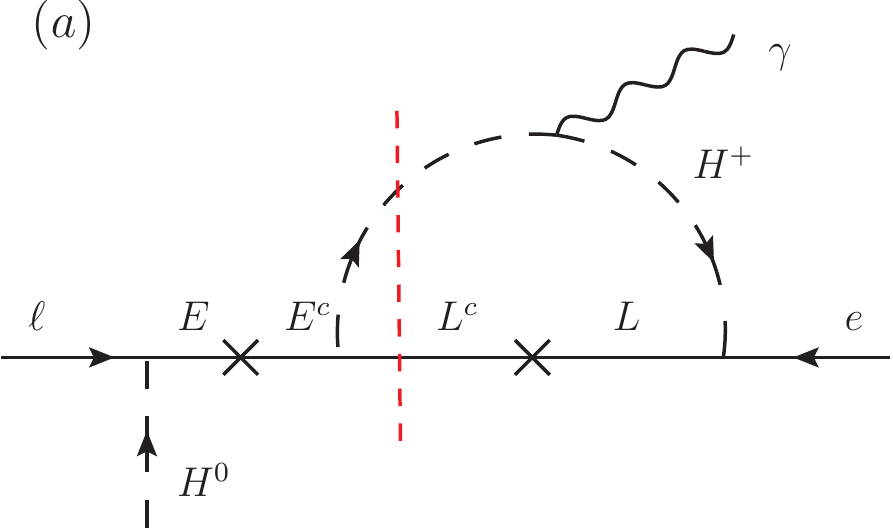}
\hspace{.5cm}
\includegraphics[width=0.3\textwidth]{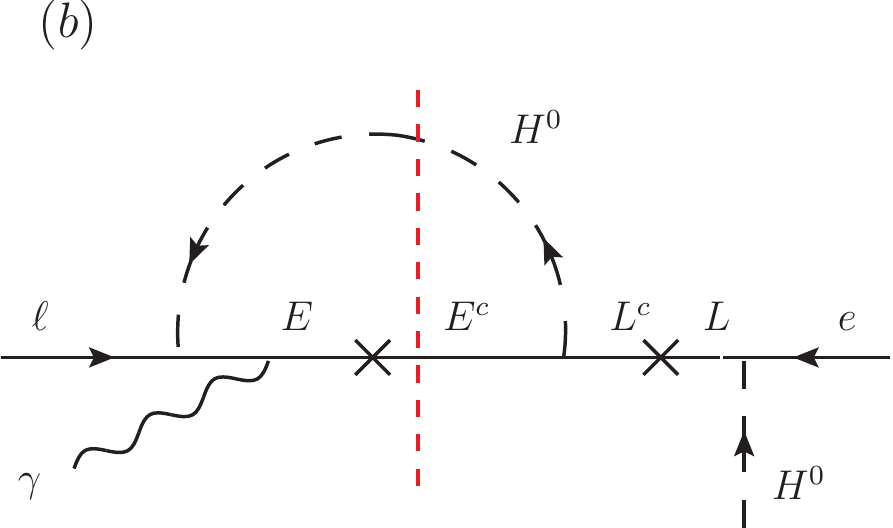}
\hspace{.5cm}
\includegraphics[width=0.3\textwidth]{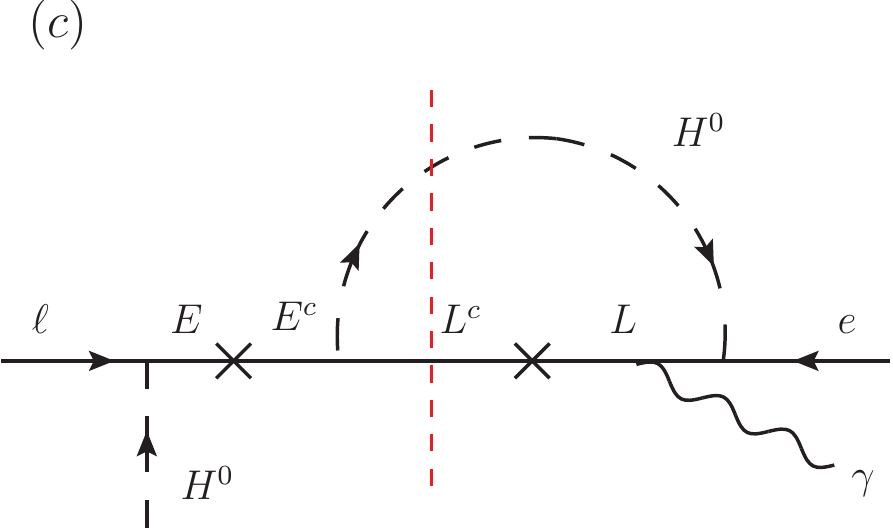}
\caption{\it 
One-loop   contributions to the $g-2$ of the SM leptons
from the model  \eq{modelEL}, with the relevant 2-cuts.
}
\label{dipoleEL}
\end{figure}

\subsection{Massive vector-like charged $E$ and doublet $L$}
\label{ELmodel}

Let us next consider a model where the extra vector-like fermions, 
$E$ and $L$, have the same  quantum numbers as the leptons of the SM. The Lagrangian  is now given by 
\be
{\cal L}=-Y_L  \ell E H-Y_R  L e H-
Y_{V}   H^\dagger L^c E^c - Y'_{V}    LE H -M_E\, E E^c- M_L\,   L L^c+h.c. 
\label{modelEL}
\ee
For the case $Y'_V\not=0$, $Y_V=0$, there is no possible Feynman diagram. Therefore we only have to study the opposite case $Y_V\not=0$, $Y'_V=0$.
The Feynman diagrams are given in Fig.~\ref{dipoleEL}.
Let us first consider the  contribution coming from the charged Higgs, diagram (a).
The calculation is identical to the one of the 2-cut of diagram (b) in Fig.~\ref{dipoleSLb}
which we already found to give
\be
\frac{\Delta C_{\gamma}}{M^2}
= \frac{Y_LY_R Y_V}{32\pi^2}\frac{1}{M_E M_L}\,.
\label{WilsoncutE}
\ee

Next we study the contributions from the neutral Higgs. For the  2-cut of diagram (b) in Fig.~\ref{dipoleEL}, the involved amplitudes read 
 \be
{\cal A}(1_\ell,3_{\gamma_-},1'_{E},3'_{H^0}) \, = \, q_e Y_L \frac{M_E}{2p_3p_{1'}}\frac{\la 33'\ra[3'{\bf1'}]}{[31]}\,, \ \ \ \
{\cal A}(3'_{\bar H^0},1'_{\bar E},2_e,4_{H^0}) \, = \, Y_RY_{V}\frac{M_L\la-{\bf1'} 2\ra}{p^2_2-M_L^2}\,.
\label{amplitudes}
\ee
This leads to
\be
{\cal A}(1_\ell,3_{\gamma_-},1'^I_{E},3'_{H^0}) \, \epsilon_{IJ} \,
{\cal A}(3'_{\bar H^0},1'^J_{\bar E},2_e,4_{H^0}) \,\simeq \,
-q_e Y_LY_R Y_{V} \frac{M_E}{M_L}
\frac{\la 32\ra}{[31]}+\cdots \, ,
\label{productE}
\ee
where we have used that $\la 33'\ra[3' 1^{'I}]\la - 1'_I 2\ra=\la 3|p_{3'}p_{1'}|2\ra= - \la 3|p_{3'}(p_3+p_1)|2\ra= - 2p_3 p_{3'} \la 32 \ra+\cdots$ $\simeq 2p_3 p_{1'} \la 32 \ra+\cdots $ with the dots corresponding to terms $\propto \la 12\ra$ which do not contribute to the dipole and to terms which are subdominant for $s_{13}/M_{L,E}^2\ll 1$.
Since \eq{productE} does not depend on the internal spinors, we can trivially integrate over the phase space and from this get
\be
\frac{\Delta C_{\gamma}}{M^2}
\, = \, -\frac{Y_LY_R Y_V}{32\pi^2}\frac{1}{M_EM_L}\,.
\label{C2contS}
\ee
As opposed to \eq{WilsoncutS} this contribution is symmetric under $E\leftrightarrow L,\ell\leftrightarrow e$. Therefore we get a factor 2 when adding the  2-cut of diagram (c) in Fig.~\ref{dipoleEL}. Summing the three contributions, we find a result in agreement with Ref.~\cite{Freitas:2014pua}.

\subsubsection{A natural zero  for models with an extra (massless) scalar singlet}
\begin{figure}[t]
\centering
\includegraphics[width=0.75\textwidth]{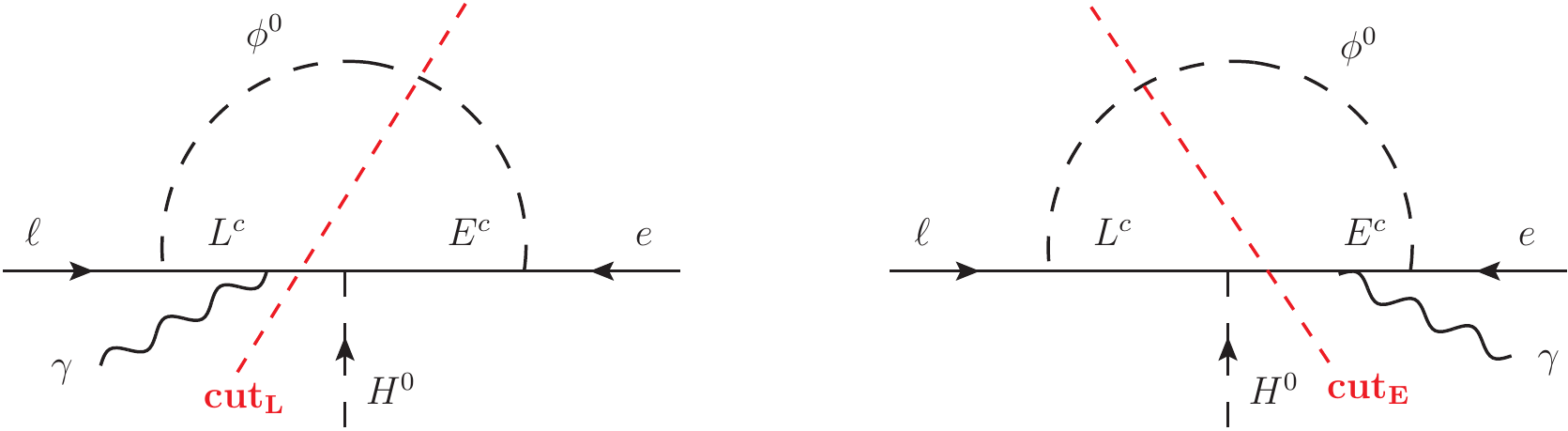}
\caption{\it 
One-loop   contributions to the $g-2$ of the SM leptons
from the model  \eq{modelphi}, with the relevant 2-cuts. 
}
\label{dipoleELphi}
\end{figure}

We have seen  that in the model \eq{modelEL} we do not find 
a vanishing contribution from the diagrams (b)+(c) since each contribution is
even under $E\leftrightarrow L,\, \ell\leftrightarrow e$.
To have a contribution which is  odd under this interchange,  we need to have the same type of diagram as the one in
 Fig.~\ref{dipoleSL} with no  mass insertions in the heavy fermion lines.
Unfortunately, diagrams of this type are identically zero
in the model \eq{modelEL} as the  Higgs line cannot be closed if we do not insert  fermion masses.
Nevertheless,  
diagrams of this type can be generated if we 
add an extra  massless scalar singlet $\phi^0$ to  the model  with the following couplings:
\be
\Delta {\cal L} \, = \, Y_L^\phi\, \phi^0\ell L^c+Y_R^\phi\, \phi^0  E^c e+h.c.
\label{modelphi}
\ee
The Feynman diagrams involving  this scalar are given in Fig.~\ref{dipoleELphi}.
Now, we can follow  the same reasoning as in  Sec.~\ref{SLmodel}
to show that this contribution to the dipole moment is zero.
Indeed, we can get the  dependence on $M_L$ of $\bf cut_{\bf E}$ 
(where $E$ is put on-shell) by noticing that it only enters in the $L$ propagator, so it must appear as $1/(M_E^2-M_L^2)$.
Dimensional analysis tells us then that $\Delta C_\gamma\propto 1/(M_E^2-M_L^2)$.
The dependence on the masses for $\bf cut_{\bf L}$ is determined by  a permutation similar to \eq{sym} with $S$ replaced by $E$ which gives 
$\Delta C_\gamma\propto 1/(M_L^2-M_E^2)$. Adding both contributions we get  zero.
It is clear that the cancellations have  nothing to do with where the photon is attached, either to  the Higgs 
line as in  Fig.~\ref{dipoleSL}  or to the fermion line as in Fig.~\ref{dipoleELphi}.

\section{$|H|^2 F^2$ Wilson coefficient}
\label{sec:GammaGammaWilsonCoefficient}

Let us now move to the calculation of the 
 Wilson coefficient of the operator contributing to the decay of a Higgs to two photons. The operator reads
\be
\frac{C_{\gamma \gamma}}{M^2} \frac{q_e^2}{2} \, |H|^2 F_{\mu \nu}^2 \,,
\ee
and the resulting amplitude is
\be
\label{HgammagammaAmplitude}
\frac{C_{\gamma \gamma} }{M^2} {\cal A}_{H^2F^2}(1_{\gamma_-},2_{\gamma_-},3_{H^0},4_{H^0})
=-\frac{C_{\gamma \gamma}}{M^2}\, q_e^2\,\la 12\ra^2\,.
\ee

We consider the same model as Eq.~\eqref{modelEL}, containing two vector-like fermions, $L$ and $E$, with the same quantum numbers as the SM leptons. Here we assume vanishing Yukawa couplings between the new fermions and the  SM leptons though, $Y_{L,R}=0$. 
In the following, we will focus on the case $Y_{V}=0, Y'_{V} \neq 0$. The discussion for the opposite case $Y_{V}\neq0, Y'_{V} = 0$ is identical. 

Since the amplitude Eq.~\eqref{HgammagammaAmplitude} does not depend on the Higgs momenta, we can take them to be zero, $p_3=p_4=0$. 
In this case we can take the limit $p_i/M\to 0$ by giving to the 
photons a  small nonzero mass $p^2\equiv p_1^2 = p_2^2 = -p_1p_2$.
An alternative is to set only one Higgs momentum to  zero, say $p_3=0$,
but in this case we have nonzero 3-cuts as we elaborate  in Appendix~\ref{appendixc}.

\begin{figure}[t]
\centering
\includegraphics[width=0.85\textwidth]{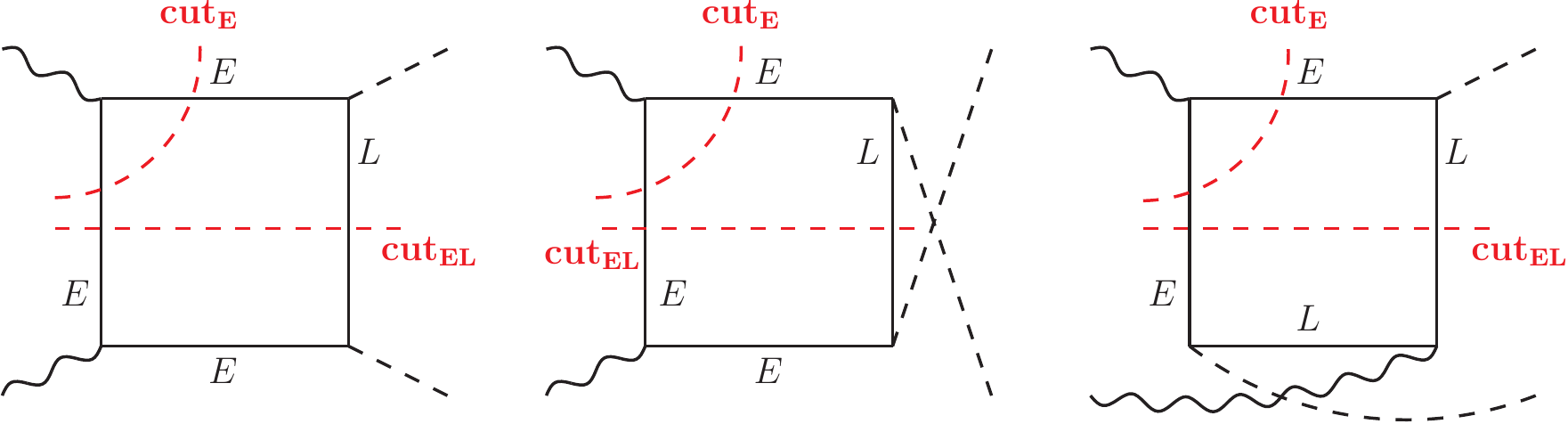}
\caption{\it One-loop   contributions to $H\gamma \gamma$ from the model \eq{modelEL}, with the 
relevant 2-cuts. There is a  similar 2-cut isolating the other photon that we do not show. Fermion lines can be clockwise  and
counterclockwise.}
\label{Higgsgammagamma}
\end{figure}

 There are three different diagrams which can contribute to the Wilson coefficient, shown in Fig.~\ref{Higgsgammagamma}. Additional contributions arise from the same diagrams with $E \leftrightarrow L$. 
 So the total contribution must be symmetric under $E \leftrightarrow L$.
 As we will see, this will clash with the fact that the contributions from Fig.~\ref{Higgsgammagamma}
are odd under $E \leftrightarrow L$.  Although to show that the total contribution is zero is quite easy, we will proceed here with  the details of the calculation which can be  useful  for cases where they do not add up to zero.

As in the $g-2$ case,  there are no possible 3-cuts or 4-cuts since the on-shell conditions cannot be simultaneously fulfilled for vanishing Higgs momenta. 
This leaves the  2-cuts shown in Fig.~\ref{Higgsgammagamma}. 
The 2-cut denoted as $\text{\bf cut}_{\bf EL}$ isolates two amplitudes involving a photon coupled to two different fermions
which are zero by gauge invariance.
This can explicitly seen by calculating these amplitudes,
\be 
\label{GammaELHAmplitude}
{\cal A}(1_{\gamma},2_E,3_L, 4_{H^0})  =  \frac{1}{s-M_E^2} \big(  \la \bm{1} | p_2 | \bm{1} ]
 \frac{\la \bm{2} \bm{3} \ra}{p} + \la \bm{1} \bm{2} \ra \la \bm{1} \bm{3} \ra \big)
  +  \frac{1}{u-M_L^2} \big( \la \bm{1} | p_3 | \bm{1} ]
   \frac{ \la \bm{3} \bm{2} \ra }{p}
  + \la \bm{1} \bm{2} \ra \la \bm{1} \bm{3} \ra \big) \, ,
\ee
where $s= (p_3 +p_4)^2$, $u= (p_2 +p_4)^2$ and  $p =\sqrt{p^2}$.
 In the limit $p_4 \rightarrow 0$, we have $s \rightarrow M_L^2$ and $u \rightarrow M_E^2$, and the amplitude vanishes after symmetrizing over the SU(2) indices of the photon.

The only remaining 2-cut is  $\text{\bf cut}_{\bf E}$ of Fig.~\ref{Higgsgammagamma}. 
This involves a   coupling of a massive photon to massive fermions given by \cite{Durieux:2019eor}
\be
\label{PhotonAmplitude}
{\cal A}(1_{\gamma},\ell_E,\ell'_E)   =  \frac{q_e}{p} \bigl(\la\bm{1} \bm{\ell} \ra \, [\bm{1} \bm{\ell'}] +   [\bm{1} \bm{\ell}] \la\bm{1} \bm{\ell'} \ra \bigr) \,.
\ee
On the other side of the cut, we have the same type of amplitude ${\cal A}(1_{\gamma},\ell_E,\ell'_E) $
but with the external fermion line corrected by the insertion of two Higgs. 
This can be considered as a  correction to the $E$ propagator which can be absorbed into 
a renormalization of the wavefunction $\delta Z_E$ and of the mass,
  \be
  M_E\to \hat M_E=M_E+\frac{1}{2} |Y'_V H^0|^2 \frac{M_E }{M_E^2-M_L^2}\,,
  \label{MEr}
  \ee
where the Higgs has  been considered  a constant configuration.
 The correction from $\delta Z_E$ is  expected to be
  exactly cancelled by a correction to the photon vertex, as dictated by gauge invariance.
This vertex correction is given by  the third diagram in Fig.~\ref{Higgsgammagamma}
where a Higgs is inserted on each of the two fermion lines. 
This leaves us with \eq{MEr} as the only effect of the Higgs.
Notice that this is odd under $E\leftrightarrow L$.

The  $\text{\bf cut}_{\bf E}$ of Fig.~\ref{Higgsgammagamma} can therefore be obtained
by the dLIPS integral of the product of two  photon couplings to the fermion $E$ of mass given by  \eq{MEr},
\be
 {\cal A}(1_{\gamma},\ell_E^I,\ell'^K_E) \, \epsilon_{IJ} \epsilon_{KL} \, {\cal A}(\ell'^L_{\bar E},\ell^J_{\bar E},2_{\gamma})
\,  = \, 
\frac{q_e^2}{p^2}
   \left[  2 \, \hat M_E^2\la \bm{1}\bm{2} \ra  [\bm{1} \bm{2} ] +  \bigl( \la \bm{1} | \ell |\bm{2} ]  \la \bm{2} | \ell' |\bm{1} ]  + (\ell \leftrightarrow \ell' )\bigr)\right],
\label{BubbleCoefficienthgammagamma}
 \ee
where we have used \eq{PhotonAmplitude}.
In order to simplify  the integration of  \eq{BubbleCoefficienthgammagamma} over dLIPS,
we can go to the rest frame of the photon where  $p_1^\mu=-p_2^\mu =(p,0,0,0) $. In this frame, we furthermore have  $\ell^\mu = (- p/2,\vec{\ell})$, where $\vec{\ell}=\ell (s_\theta c_\phi,  s_\theta s_\phi,   c_\theta)$ with $\ell =(p^2/4-M_E^2)^{1/2}$, and  $\ell'^\mu=-\ell^\mu-p_1^\mu$. We next have the freedom to choose a particular basis for the photon polarizations, for which we take (equivalent to setting $|s|=\sin(\theta/2)=0$ in Appendix C of Ref.~\cite{Arkani-Hamed:2017jhn}):
\be
\label{spinors12} 
| 1^{I=1} \ra = \sqrt{ p} \begin{pmatrix} 1 \\ 0 \end{pmatrix}\, ,  \qquad \quad | 1^{I=2} \ra = \sqrt{p} \begin{pmatrix} 0 \\ 1 \end{pmatrix} \,,
\ee
and $| 1^{I=1,2} ]  = | 1^{I=1,2}\ra $. Since $p_1=-p_2$, we can set $|\bf 2\ra = -|\bf 1\ra$ and $|\bf 2]= |\bf 1]$. Furthermore, we will only calculate the result for  one of the three polarizations of the massive photon, as the others must give the same result.
 We take the longitudinal one, which leads to 
\be 
\la \bm{1} \bm{2}\ra  [\bm{1} \bm{2}]\, \rightarrow \, \la 1^{I=1} 2^{I=2} \ra  [1^{I=1} 2^{I=2}] \, = \, p^2 ,
\ee
and similarly for the other terms in Eq.~\eqref{BubbleCoefficienthgammagamma}.
This gives
\be
\begin{split}
\int d\text{LIPS}\,  {\cal A}(1_{\gamma},\ell_E,\ell'_E) \times {\cal A}(\ell'_{\bar E},\ell_{\bar E},2_{\gamma})
& \, \rightarrow \, 
q_e^2\int_0^{2\pi}\!\!\frac{d \phi}{4\pi} \int_0^\pi \!\! d \theta\, s_\theta\Big[ 2 \, \hat M^2_E ( 1-c^2_\theta)+
\frac{1}{2}p^2(1+c^2_\theta)
\Big]\\ 
&\, = \, \frac{2q_e^2}{3}\left(2\hat M^2_E+ p^2\right).
\label{product}
\end{split}
\ee
Furthermore, from the bubble integral we have expanding in $p^2/M^2$ and $|H^0|^2/M^2$ that
\be
I_2(p^2,\hat M_E^2,\hat M_E^2) = \frac{1}{16 \pi^2} \left( \frac{1}{\epsilon} +\ln \frac{\mu^2}{\hat M_E^2} 
+ \dots\right)=  \frac{1}{16 \pi^2} \left(
 \frac{1}{\epsilon} +\ln \frac{\mu^2}{ M_E^2} -
\frac{ |Y'_V H^0|^2  }{M_E^2-M_L^2}+\cdots\right).
\label{I2E}
\ee
In order to match the product of 
\eq{product} and \eq{I2E} with the amplitude Eq.~\eqref{HgammagammaAmplitude}, we must take the photons in the latter to be massive too, $\la 12\ra^2\to \la {\bf 12}\ra^2$, and  project the photons into their longitudinal components, $\la {\bf 12}\ra^2 \to p^2$.
The Wilson coefficient then arises from the $p^2|H^0|^2$-dependent part of this product which gives\footnote{The  leading terms (proportional to $\hat M^2_E$ in \eq{product}) must vanish when adding other bubbles which do not involve $p^2$ terms. These have been omitted here such as 
the one related to the vertical 2-cut  of  the diagrams of Fig.~\ref{Higgsgammagamma} which isolates a two-photon amplitude.} 
\be
\frac{\Delta C_{\gamma \gamma}}{M^2}  \, = \, \frac{1}{16 \pi^2} \frac{2}{3} \frac{|Y_V'|^2}{M_E^2-M_L^2} \, .
\label{resultHgg}
\ee
This is again odd under the interchange $E\leftrightarrow L$, so when adding the contribution 
from the same diagrams as in Fig.~\ref{Higgsgammagamma} but with $E\leftrightarrow L$,
we see that the total contribution to the Wilson coefficient vanishes.

\section{Conclusions}

In this paper we have shown how to efficiently calculate finite contributions to Wilson coefficients using amplitude methods which are known to significantly simplify  loop calculations   compared to the Feynman approach. 
The Wilson coefficients can be extracted from one-loop amplitudes by expanding them in powers of the masses of the heavy particles.
Using a Passarino-Veltman decomposition, one-loop amplitudes can in turn be expressed in terms of basic scalar integrals called  bubbles, triangles and boxes. 
By applying generalized unitarity cuts to this relation, the coefficients of these integrals are obtained from products of tree-level amplitudes integrated over some phase-space. In general, the one-loop amplitudes receive additional contributions from rational terms whose calculation is more involved. We have shown, however, that these rational terms cannot contribute to the Wilson coefficients. Combining everything, the Wilson coefficients can then be calculated from products of, in general, two, three and four on-shell amplitudes.

We have applied this method to calculate  finite contributions
to   the dipole-moment operator and the operator coupling the Higgs to photons, $|H|^2 F^2$. 
This was done for several theories  containing  heavy vector-like fermions. 
We have shown that the  calculation simplifies significantly by taking the momenta of the (one respectively two) Higgs fields in these operators to zero. 
The reason is that in this limit, triangles and boxes in the Passarino-Veltman decomposition necessarily vanish due to kinematical constraints. 
The calculation of the Wilson coefficients then boils down to products of two on-shell amplitudes, integrated over the phase space of the two intermediate particles. 
In many cases, this phase-space integral is trivial, further simplifying the calculation.

Our method has allowed to  shine light on  the mysterious  cancellations  in the contributions to  these Wilson coefficients,
 recently discussed in detail in \cite{Arkani-Hamed:2021xlp} for the dipole-moment operator, 
but  also  noted (e.g. in \cite{Panico:2018hal}) for $|H|^2F^2$.
This has been shown to happen in 
certain models with heavy fermions, even though the contributions do not   seem to be forbidden by any symmetry. 
We have seen that the Wilson coefficients can be expressed as the sum of two different products of amplitudes, corresponding to two possible ways of applying 2-cuts to the one-loop amplitude. 
We have found that these contributions to the Wilson coefficients are odd under the exchange of the heavy fermions, while 
the total contribution has to be even under this exchange. 
Therefore, upon inspection with amplitude methods, the cancellation  boils down to a clash of $even\times odd$ under the exchange parity. 
This understanding  has allowed us to find other models where this cancellation also occurs.

\medskip
\section*{Acknowledgments}
We thank Pietro Baratella for collaboration at early stages of this project, 
and Roberto Pittau for useful discussions on rational terms.
A.P.~has been  supported by the Catalan ICREA Academia Program, and  the grants  2014-SGR-1450, PID2020-115845GB-I00/AEI/10.13039/501100011033.
L.D.R. has been supported by a fellowship from "la Caixa" Foundation (ID 100010434) and from the European Union's Horizon 2020 research and innovation programme under the Marie
Sklodowska-Curie Action grant agreement No 847648.

\appendix

\section{The absence of rational terms}
\label{appendixa}

Here we will show that the rational terms $R$ of \eq{general}  do not contribute to 
the Wilson coefficients arising from integrating out heavy particles.

Rational terms are related to UV divergences and, in $D=4-2\epsilon$ dimensions, within the Passarino-Veltman reduction, they appear from the product $\epsilon I_1$ or $\epsilon I_2$, where the scalar integrals $I_1$ and $I_2$ 
carry  $1/\epsilon$ UV-divergent terms. 
Having no imaginary parts, they can in principle not be obtained by performing cuts of the loop diagrams. 
Nevertheless, it was  shown  in  \cite{Badger:2008cm}
that by extending \eq{general} to 
a $D$-dimensional Passarino-Veltman decomposition, generalized unitarity methods can also be used to obtain these rational terms.

Let us here review the argument of \cite{Badger:2008cm}. This is based on the observation that, by promoting the loop integration momentum $l$
to $D=4-2\epsilon$ dimensions, any rational term can only appear from the $-2\epsilon$ component of $l^2$, namely $l^2 = l_{(4)}^2 + l^2_{(-2\epsilon)} \equiv l_{(4)}^2 - \mu^2$.
The usual basis of the Passarino-Veltman decomposition of \eq{general} is then enlarged with new master integrals whose integrands contain powers of $\mu^2$.\footnote{In \cite{Badger:2008cm}, the new master integrals are denoted as $I_n^{4-2\epsilon}[\mu^{2k}]$ where $\mu^{2k}$ is understood as being integrated over. The explicit computation of these integrals gives rise to the  coefficients  in \eq{eq:rational}.}
Within this framework, the rational terms can be obtained by exploiting generalized unitarity methods and can be written as 
\cite{Badger:2008cm}
\be
\label{eq:rational}
R  \, = \, -\frac{1}{6} \sum_{i,j} \tilde C_2^{(ij)} (s_{ij} - 3(M_i^2 + M_j^2)) - \frac{1}{2} \sum_b \tilde C_3^{(b)}   - \frac{1}{6}  \sum_c \tilde C_4^{(c)}\,.
\ee
Without going into the details of the proof of the above formula (we refer the interested reader to Ref.~\cite{Badger:2008cm}), it will be enough to highlight how the coefficients $\tilde C_n$ can be extracted. 
Firstly, since the extra component $\mu^2$ in the integration momentum $l^2$ can be effectively seen as a mass, 
 $l_{(4)}^2 = M_i^2 + \mu^2$,   all internal masses in the one-loop amplitude must be  
 shifted by the same mass parameter $\mu^2$, i.e.~$M_i^2 \to M_i^2 + \mu^2$.
Secondly, we must perform the corresponding $n$-cuts of the amplitude,  and take the limit of large $\mu^2$;
  $\tilde C_2$ and $\tilde C_3$ are given  from the coefficient of the $\mu^2$-term of this expansion, while $\tilde C_4$ is obtained from the $\mu^4$-term.

Equipped with these observations, we are ready to show the absence of rational terms in the Wilson coefficients considered in this work. 
We are interested in one-loop contributions arising from  renormalizable theories in the limit  $M\gg p_i$,
where we match to the EFT at order $1/M^2$.
Therefore the loop integrals must converge to zero for large $M$.
Nevertheless, since the rational terms are obtained in the large-$\mu^2$ limit ($\mu^2 \gg M^2, s_{ij}$),  $M^2$ is always subleading with respect to $\mu^2$. Indeed, the  internal propagators will be given by
 $1/({s_{ij} - M^2 - \mu^2})$ and $M^2$ can only appear as powers of $M^2/\mu^2$ in the  $\mu^2 \to \infty$ limit, and therefore cannot contribute at  $O(1/M^2)$.\footnote{Rational terms can appear at $O(1/s_{ij})$. However, these singular contributions are cancelled by  contributions from $C_nI_n$ to make the one-loop amplitude  non-singular in the limit of small momenta.}

\section{Massless and massive spinor-helicity variables}
\label{appendixb}

We begin with specifying our conventions for massless and massive spinor-helicity variables. We choose the metric $\eta_{\mu \nu} = \text{diag}(+,-,-,-)$. For a massless particle, the momentum can be written as 
\be
p_{\alpha \dot{\alpha}} \, = \, |p \ra_\alpha  [p|_{\dot{\alpha}} \, ,
\ee
where $| p \ra_\alpha$ and $ [p|_{\dot{\alpha}}$ are two-component spinors which transform under the little group with helicity $h=\mp 1/2$, respectively. The Lorentz indices $\alpha$ and $\dot{\alpha}$ are raised and lowered with the two-component Levi-Civita symbol, which we choose such that $\epsilon^{12} = -\epsilon_{12}=1$. In particular, we have
\be
\la p|^\beta \, = \, \epsilon^{\beta \alpha} | p \ra_\alpha \, , \qquad | p ]^{\dot{\alpha}} \, = \, \epsilon^{\dot{\alpha} \dot{\beta}}[p|_{\dot{\beta}}  \, .
\ee
The Lorentz indices of two angle or square brackets are contracted as
\be
\la p q \ra \, = \, \la p|^\alpha \, | p \ra_\alpha\,, \qquad [pq]\, = \, [p|_{\dot{\alpha}} \, | p ]^{\dot{\alpha}} \, .
\ee

For a massive particle, on the other hand, we need twice as many spinors which combine into two vectors transforming under the little group SU(2), $| p \ra_\alpha^I$ and $| p ]^{\dot{\alpha} \, I}$ \cite{Arkani-Hamed:2017jhn}. The SU(2) little-group indices $I$ are again raised and lowered with the two-component Levi-Civita symbol. The momentum is then given by
\be 
p_{\alpha \dot{\alpha}} \, = \, \epsilon_{IJ} | p \ra_\alpha^I  [p|_{\dot{\alpha}}^J \, = \, | p \ra_\alpha^I  [p|_{\dot{\alpha}\, I} \, .
\ee
The massive spinor-helicity variables fulfil the identities
\be
\begin{aligned}
\label{BoldVariableIdentities}
\la p^I p^J \ra \, & = \, - M \epsilon^{IJ}, & \qquad   [ p_I p_J ]  & = \, - M \epsilon_{IJ} \, ,  \\
| p \ra_\alpha^I  \,\la p|^{\beta}_I  \, & = \, - M \delta_\alpha^\beta \, , & \qquad  | p ]^{\dot{\alpha} \, I}  \, [p|_{\dot{\beta}\, I} & = \, M \delta^{\dot{\alpha}}_{\dot{\beta}} \, ,
\end{aligned}
\ee
where $M$ is the mass of the particle. From this, we in particular obtain the Dirac equation
\be
p | p]^I \, = \, M |p\ra^I,  \qquad \, \, p | p\ra^I \, = \, M |p]^I \, .
\ee
We will often not write the SU(2) indices explicitly and will then use bold letters for the massive spinor-helicity variables, $| p \ra^I = | \bm{p} \ra$ and $| p ]^I = | \bm{p} ]$, to distinguish them from the massless ones.
In  amplitudes involving massive spinors, the SU(2) indices  $I,J,...$ of a given state
must be symmetrized \cite{Arkani-Hamed:2017jhn}.

When contracting amplitudes to obtain the coefficients of the Pasarino-Veltman decomposition, we need to flip the momenta of the particles on one side of the contraction from incoming to outgoing (amplitudes are defined with all momenta being incoming). The spinor-helicity variables for the flipped momenta then satisfy different contraction rules from the ones given above \cite{Durieux:2019eor}:
\be
\begin{aligned}
\label{BoldVariableIdentities}
 | p \ra_\alpha^I  [- p|_{\dot{\alpha}\, I} \, & = \, p_{\alpha \dot{\alpha}} \, , & \qquad  | p ]^{\dot{\alpha} \, I} \la - p|^{\alpha}_I \, & = \, p^{ \dot{\alpha} \alpha} \\
| p \ra_\alpha^I  \,\la - p|^{\beta}_I  \, & = \, M \delta_\alpha^\beta \, , & \qquad  | p ]^{\dot{\alpha} I}  \, [- p|_{\dot{\beta}I} \, & = \, M \delta^{\dot{\alpha}}_{\dot{\beta}} \, .
\end{aligned}
\ee

\section{An alternative way to calculate the $|H|^2 F^2$ Wilson coefficient}
\label{appendixc}

In the following, we will discuss an alternative way to calculate the Wilson coefficient for the operator $|H|^2 F^2$, corresponding to the amplitude in Eq.~\eqref{HgammagammaAmplitude}. Here we keep the photons on-shell and take the momentum of only one Higgs to be identically zero, $p_3=0$. Since then $p_4^2= 2 p_1 p_2$, the other Higgs must be slightly off-shell. The diagrams which contribute to the Wilson coefficient are those shown in Fig.~\ref{Higgsgammagamma} plus the same diagrams with $E \leftrightarrow L$. We will see, however, that in the chosen kinematical configuration the nonvanishing cuts are different from those depicted in Fig.~\ref{Higgsgammagamma}. Let us first consider 3-cuts and 4-cuts.
As before, cutting both fermion lines attached to the Higgs with vanishing momentum gives zero since the on-shell conditions cannot be simultaneously fulfilled. This eliminates all 4-cuts and several of the 3-cuts. Furthermore, we have shown in Sec.~\ref{sec:GammaGammaWilsonCoefficient} that the amplitude in Eq.~\eqref{GammaELHAmplitude} for a massive photon vanishes identically for zero Higgs momentum. Since this amplitude contains the corresponding amplitudes for a massless photon (taking the high-energy limit), the latter are zero too. Any 3-cut which isolates this amplitude therefore gives no contribution. 
We are then left with only one 3-cut, the one that  puts the three fermions $E$ in the first two diagrams in Fig.~\ref{Higgsgammagamma} on-shell (plus the corresponding 3-cut in the diagrams with $E \leftrightarrow L$).

Let us now calculate this 3-cut. We denote the 
momentum of the fermion line connecting the two photons as $\ell$. One solution for this momentum after restricting three of the propagators in the loop to be on-shell can be parametrized as \cite{Forde:2007mi}
\be
\label{CutMomentum}
\ell \, = \, \tau \,  |1 \ra [2| \, - \, \tau^{-1}  \frac{M_E^2}{s_{12}} \, | 2 \ra [1| \, ,
\ee
where $\tau$ is the remaining integration variable. The other two cut momenta are $\widetilde{\ell}= -\ell-p_1$ and $\widehat{\ell}= \ell-p_2$, altogether satisfying $\ell^2=\widetilde{\ell
}^2=\widehat{\ell}^2=M_E^2$ as required. The other solution $\ell^*$ is given by Eq.~\eqref{CutMomentum} with $1 \leftrightarrow 2$. The triangle coefficient then is\footnote{In general, one has to expand the integrand for large $\tau$. This is not necessary in our case.}
\be
C_3\, = \,  \sum_{\ell, \ell^*}
\int d\tau J_\tau \, {\cal A}(1_{\gamma_-},\widetilde{\ell}_E,\ell_E)\times {\cal A}(2_{\gamma_-},\ell_{\bar E},\widehat{\ell}_{\bar E}) \times {\cal A}(\widetilde{\ell}_{\bar E},\widehat{\ell}_{E},3_{H^0},4_{H^0})\,,
\label{C3}
\ee
where $J_\tau$ is a Jacobian arising from the transformation to the integration variable $\tau$. 
We have $\int d\tau J_\tau \tau^n = \delta_{0n}$ which leads to
\be
\label{C3HiggsGammaGamma}
C_3 \, = \, \sum_{\ell, \ell^*}
\bigl[ {\cal A}(1_{\gamma_-},\widetilde{\ell}_E,\ell_E)\times {\cal A}(2_{\gamma_-},\ell_{\bar E},\widehat{\ell}_{\bar E}) \times {\cal A}(\widetilde{\ell}_{\bar E},\widehat{\ell}_{E},3_{H^0},4_{H^0})\bigr]_{\tau^0}\, .
\ee

The 4-point amplitude which is isolated in the 3-cut gets contributions from both the first and second diagram in Fig.~\ref{Higgsgammagamma} and reads
\be
\label{TwoPhotonsTwoHiggsAmplitude}
 {\cal A}(\widetilde{\ell}_{\bar E},\widehat{\ell}_{E},3_{H^0},4_{H^0}) \, = \, |Y_V'|^2 \left(\frac{[ -\bm{\widetilde{\ell}}\, |(\widehat{\ell}+p_3)|\bm{\widehat{\ell}}\, \ra}{(\widehat{\ell}+p_3)^2 -M_L^2} \, - \, \frac{[ \bm{\widehat{\ell}}\,|(-\widetilde{\ell}+p_3)|-\bm{\widetilde{\ell}}\, \ra}{(-\widetilde{\ell}+p_3)^2- M_L^2}\right) \, \underset{p_3 \rightarrow 0}{=} \, 2 |Y_V'|^2 \frac{ M_E  \, [ -\bm{\widetilde{\ell}}\, \bm{\widehat{\ell}}\, ]}{M_E^2-M_L^2} \, .
\ee
In the second step we have taken the Higgs momentum $p_3$ to zero. Furthermore, the 3-point amplitudes are given by
\be
{\cal A}(1_{\gamma_-},\widetilde{\ell}_E,\ell_E) \, = \, q_e \frac{\la 1|  \ell | \xi]}{M_E [1 \xi]} \, [ \bm{\widetilde{\ell}} \bm{\ell} ]\,, 
\ee
with $\xi$ being a reference spinor, and similarly for ${\cal A}(2_{\gamma_-},\ell_{\bar E},\widehat{\ell}_{\bar E}) $. Using Eq.~\eqref{CutMomentum}, the reference spinors drop out. Combining the three amplitudes and summing over the SU(2) indices of the internal fermions, Eq.~\eqref{C3HiggsGammaGamma} then yields
\be
\label{TripleCut}
C_3 \, = \,   \frac{16 \, |Y_V'|^2}{M_E^2-M_L^2} \,\frac{M_E^4} {s_{12}} \, q_e^2 \la 1 2 \ra^2 \, .
\ee
Here we have included a factor 2 due to the fact that the fermion lines in the loop can be clockwise or counterclockwise (or, alternatively, that the photons can be interchanged, $1\leftrightarrow 2$).
From Eq.~\eqref{general}, the contribution to the one-loop amplitude ${\cal A}_{H^2F^2}$ is obtained by multiplying Eq.~\eqref{TripleCut} with the triangle function corresponding to this 3-cut,
\be
I_3(s_{12},0,0; M_E,M_E,M_E) \, = \, \frac{1}{16 \pi^2} \left( \frac{1}{2 M_E^2} \, + \, \frac{s_{12}}{24 M_E^4} \, + \dots \right), 
\ee
with $s_{12} = (p_1+p_2)^2$ and which we have expanded to $O(s_{12}/M_E^2)$. The leading term in this expansion gives rise to an unphysical pole term when multiplied with Eq.~\eqref{TripleCut}. This is cancelled by a rational term. Using Eq.~\eqref{master}, the next term in the expansion then yields the contribution of the 3-cut to the Wilson coefficient
\be
\label{DeltaCHiggsGammaGamma}
\frac{\Delta C_{\gamma \gamma} }{M^2} \, = \, \frac{1}{16\pi^2}  \frac{2}{3} \frac{|Y_V'|^2}{M_E^2-M_L^2} \, ,
\ee
in agreement with Eq.~\eqref{resultHgg}. This is again odd under the exchange $E \leftrightarrow L$ and is therefore exactly cancelled by the contribution which arises from the diagrams in Fig.~\ref{Higgsgammagamma} with $E \leftrightarrow L$. We thus conclude that neither 4-cuts nor 3-cuts contribute to the Wilson coefficient $C_{\gamma \gamma}$.

We still need to discuss the 2-cuts. Since the 3-cut already accounts for the result in Eq.~\eqref{resultHgg}, we expect that the 2-cuts give no additional contributions to the Wilson coefficient. As before, the 2-cut which leaves the zero-momentum Higgs alone vanishes since the on-shell conditions cannot be fulfilled. Furthermore, any 2-cut which isolates an on-shell photon (with $p^2=0$) 
does not contribute to the Wilson coefficient since the corresponding bubble integrals do not depend on any momentum (cf.~the discussion in Sec.~\ref{SLmodel}). Other 2-cuts are zero since they again cut out an amplitude of the type ${\cal A}(1_{\gamma_-},3_{H^0},\ell_E,\ell'_L)$ which vanishes for $p_3=0$. Finally, also the 2-cut which leaves the off-shell Higgs alone does not contribute. Indeed, the amplitude on the other side of this cut is ${\cal A}(1_{\gamma_-}, 2_{\gamma_-},3_{H^0},\ell_E,\ell'_L)$ whose only kinematically allowed factorization channels once again involve amplitudes of the type ${\cal A}(1_{\gamma_-},3_{H^0},\ell_E,\ell'_L)$ with $p_3=0$. The amplitude ${\cal A}(1_{\gamma_-}, 2_{\gamma_-},3_{H^0},\ell_E,\ell'_L)$ and the corresponding 2-cut therefore vanish too.
This leaves only the 2-cut which puts the upper and lower $E$ line in the first two diagrams in Fig.~\ref{Higgsgammagamma} on-shell. Notice that the same two lines are also cut in the non-vanishing 3-cut. The 2-cut therefore obtains a contribution from the corresponding triangle and is guaranteed to be non-zero. It can be shown that the 2-cut exactly matches the latter and the Wilson coefficient therefore does not get any additional contribution from this 2-cut either.


\begin{thebibliography}{99}


\bibitem{Arkani-Hamed:2021xlp}
N.~Arkani-Hamed and K.~Harigaya,
JHEP \textbf{09} (2021), 025
[arXiv:2106.01373 [hep-ph]].


\bibitem{Panico:2018hal}
G.~Panico, A.~Pomarol and M.~Riembau,
JHEP \textbf{04} (2019), 090
[arXiv:1810.09413 [hep-ph]].

\bibitem{Cheung:2015aba}
C.~Cheung and C.~H.~Shen,
Phys. Rev. Lett. \textbf{115} (2015) no.7, 071601
[arXiv:1505.01844 [hep-ph]].



\bibitem{Shadmi:2018xan}
Y.~Shadmi and Y.~Weiss,
JHEP \textbf{02} (2019), 165
[arXiv:1809.09644 [hep-ph]].


\bibitem{Bern:2021ppb}
Z.~Bern, D.~Kosmopoulos and A.~Zhiboedov,
J. Phys. A \textbf{54} (2021) no.34, 344002
[arXiv:2103.12728 [hep-th]].

\bibitem{Craig:2021ksw}
N.~Craig, I.~G.~Garcia, A.~Vainshtein and Z.~Zhang,
[arXiv:2112.05770 [hep-ph]].

\bibitem{conf}
Part of this work was presented by A.~Pomarol at the workshop ``Portoro\v z 2021. Physics of the flavourful Universe'',  September 2021, Portoro\v z, Slovenia; http://hepworkshop.ijs.si/2021.



\bibitem{Dixon:2013uaa}
L.~J.~Dixon,
``A brief introduction to modern amplitude methods,''
[arXiv:1310.5353 [hep-ph]].


\bibitem{Caron-Huot:2016cwu}
S.~Caron-Huot and M.~Wilhelm,
JHEP \textbf{12} (2016), 010
[arXiv:1607.06448 [hep-th]].

\bibitem{2}
J.~Elias Mir\'o, J.~Ingoldby and M.~Riembau,
JHEP \textbf{09} (2020), 163
[arXiv:2005.06983 [hep-ph]].

\bibitem{Baratella:2020lzz}
P.~Baratella, C.~Fernandez and A.~Pomarol,
Nucl. Phys. B \textbf{959} (2020), 115155
[arXiv:2005.07129 [hep-ph]].

\bibitem{4}
M.~Jiang, T.~Ma and J.~Shu,
JHEP \textbf{01} (2021), 101
[arXiv:2005.10261 [hep-ph]].

\bibitem{5}
P.~Baratella, C.~Fernandez, B.~von Harling and A.~Pomarol,
JHEP \textbf{03} (2021), 287
[arXiv:2010.13809 [hep-ph]].

\bibitem{6}
Z.~Bern, J.~Parra-Martinez and E.~Sawyer,
JHEP \textbf{10} (2020), 211
[arXiv:2005.12917 [hep-ph]].

\bibitem{7}
P.~Baratella, D.~Haslehner, M.~Ruhdorfer, J.~Serra and A.~Weiler,
[arXiv:2109.06191 [hep-th]].

\bibitem{8}
M.~Accettulli Huber and S.~De Angelis,
JHEP \textbf{11} (2021), 221
[arXiv:2108.03669 [hep-th]].

\bibitem{Arkani-Hamed:2017jhn}
N.~Arkani-Hamed, T.~C.~Huang and Y.~t.~Huang,
JHEP \textbf{11} (2021), 070
[arXiv:1709.04891 [hep-th]].



\bibitem{Freitas:2014pua}
A.~Freitas, J.~Lykken, S.~Kell and S.~Westhoff,
JHEP \textbf{05} (2014), 145
[erratum: JHEP \textbf{09} (2014), 155]
[arXiv:1402.7065 [hep-ph]].

\bibitem{Durieux:2019eor}
G.~Durieux, T.~Kitahara, Y.~Shadmi and Y.~Weiss,
JHEP \textbf{01} (2020), 119
[arXiv:1909.10551 [hep-ph]].

\bibitem{Badger:2008cm}
S.~D.~Badger,
``Direct Extraction Of One Loop Rational Terms,''
JHEP \textbf{01} (2009), 049
[arXiv:0806.4600 [hep-ph]].




\bibitem{Forde:2007mi}
D.~Forde,
Phys. Rev. D \textbf{75} (2007), 125019
[arXiv:0704.1835 [hep-ph]].


\end{thebibliography}
\end{document}